\documentclass[preprint2]{aastex}

\newcommand{\gapprox}{\lower.4ex\hbox{$\;\buildrel >\over{\scriptstyle\sim}\;$}}
\newcommand{\lapprox}{\lower.4ex\hbox{$\;\buildrel <\over{\scriptstyle\sim}\;$}}
\newcommand{\begeq}{\begin{equation}}
\newcommand{\fineq}{\end{equation}}

\def\ellprime0{\ell'_0}

\slugcomment{accepted, ApJ}

\shorttitle{CME-driving currents}

\shortauthors{Subramanian, \& Vourlidas}

\begin{document}

\title{DRIVING CURRENTS FOR FLUX ROPE CORONAL MASS EJECTIONS}

\author{Prasad Subramanian\altaffilmark{1}}
\affil{Indian Institute of Science Education and Research, Sai Trinity Building, Pashan, Pune - 411021, India}
\and
\author{Angelos Vourlidas}
\affil{Code 7663, Naval Research Laboratory, Washington, DC 20375, USA}

\vfil

\altaffiltext{1}{p.subramanian@iiserpune.ac.in}

\begin{abstract}
  We present a method for measuring electrical currents enclosed by
  flux rope structures that are ejected within solar coronal mass
  ejections (CMEs). Such currents are responsible for providing the
  Lorentz self-force that propels CMEs. Our estimates for the driving
  current are based on measurements of the propelling force obtained
  using data from the LASCO coronagraphs aboard the SOHO satellite. We
  find that upper limits on the currents enclosed by CMEs are
  typically around $10^{10}$ Amperes. We estimate that the magnetic
  flux enclosed by the CMEs in the LASCO field of view is a few$\times
  10^{21}$ Mx.
\end{abstract}



\section{Introduction}

Although it is well known that the dynamics of magnetic fields in the
solar corona control most key aspects of its physical state, there are
only a few known estimates of currents therein. They range from
currents in solar prominences inferred from their dynamics (e.g.,
Ballester \& Kleczek 1984), indirect estimates of currents in flaring
loops (e.g., Zaitsev et al 1998; Tan et al 2006) and estimates of
currents inferred from differential Faraday rotation measures of
background sources observed against the corona (Spangler 2007). The
Faraday rotation technique has been used to infer the magnetic field
direction in CMEs (Liu et al 2007) but the magnitude of the magnetic
field (or the associated currents) cannot be determined.  In this
work, we demonstrate a new method of estimating coronal currents;
specifically, the driving currents enclosed by flux rope CMEs.  To the
best of our knowledge, this is the first direct estimate of the
current enclosed by CMEs in the LASCO C2 and C3 field of view.

Vourlidas et al (2000) and Subramanian \& Vourlidas (2007) (hereafter, SV07) have shown that the dissipation of 
energy contained in the magnetic fields entrained by CMEs is a viable means of powering them
in the $\sim$ 2--30 $R_{\odot}$ field of view. These conclusions were arrived at by examining the 
requirements on the mechanical driving power from LASCO data, and by deriving reasonable
upper and lower limits on the magnetic field carried by the CMEs. This work addressed
the overall energetics of the problem without going into the details of how the magnetic
energy is utilized in driving the CME. 

We re-examine the flux-rope CME sample
considered by SV07. This is the definitive sample of flux rope CMEs
observed by the LASCO instrument between 1996 and 2001, prepared after
careful inspection of the data in order to confirm their flux rope morphology. Furthermore, SV07 have chosen CMEs that retain a clear flux
rope morphology throughout their propagation in the LASCO C2 and C3
fields of view. This means that these CMEs largely remain in the plane
of the sky through the duration of these observations and projection
effects (which are sensitive functions of the inclination from the
plane of the sky; e.g., Vourlidas \& Howard 2006) can be considered to be minimal. Of the full flux rope CME sample, we consider only the CMEs in group A of SV07; the ones whose
mechanical (i.e., kinetic + potential) energy increases with time in the LASCO C2 and C3 fields
of view. This means that these CMEs experience a driving force in this height range. In other words, we use the best sample of {\em driven, flux rope} CMEs in
this paper.

\section{Procedure and Results}
\subsection{Forces on the CME loop}
The outward force $f$ per unit length on the flux rope (in cgs units) is given by

\begin{eqnarray}
\nonumber
f = \frac{I_{0}^{2}}{c^{2}\,a}\,\biggl [ {\rm ln}\biggl (\frac{8\,a}{r_{0}}\biggr ) + \frac{l_{i}}{2} - \frac{3}{2} \biggr ] + \pi\,r_{0}^{2} \bigl ( \vec{\nabla}\, P_{\infty} \bigr ) + \\
\vec{I} \times \vec{B}_{\infty} - \pi\,\rho\,r_{0}^{2}\,\frac{G\,M_{\odot}}{a^{2}}\, ,
\label{eq1}
\end{eqnarray}

This expression is similar to the one employed by Yeh (1995), and includes
all the possible forces on the flux rope. The first term on the right hand side (RHS) is the Lorentz self-force; this was not included in Yeh's (1995) treatment, for he was concentrating on a regime where the curvature of the flux rope was negligibly small. Physically, the Lorentz self-force can be understood as follows: 
consider a section of a current-carrying tube of cross-sectional radius $r_{0}$, where the current flows axially, generating a toroidal magnetic field. When this tube is bent into a circular section of
radius $a$, one can envisage the toroidal field lines ``bunching up'' at the bottom of the tube and
getting sparse on the top. This means that magnetic pressure associated with the toroidal field lines
will be greater at the bottom than at the top. This gradient in pressure results in an outwardly directed
force on the bent tube (e.g., Mouschovias \& Poland 1978). 
We will discuss this term in further detail shortly. 
The second term expresses the effect of buoyancy owing to the gradient of the ambient pressure ($P_{\infty}$) that drives the solar wind, and $\pi r_{0}^{2}$ represents the cross sectional area of the CME. The third term on the RHS is due to the Lorentz forces between the current $\vec{I}$ carried by the flux rope CME and external magnetic fields $\vec{B}_{\infty}$. The fourth term arises from the gravitational pull of the sun on the CME, where $\rho$ is the matter density inside the CME and $a$ is the distance of its center of mass from the sun center. One salient feature of terms 2 and 4 is the fact that they are both proportional to the cross-sectional area of the CME. SV07 have established that the mechanical force acting on the CME is independent of its cross-sectional area. This result, which was one of the salient ones in SV07, is certainly valid for the subset of the SV07
sample we study here, and is probably of more general validity. With regard
to the present study, this implies that terms 2 and 4 on the RHS of Eq~\ref{eq1} (which are proportional to the CME cross-sectional area) are unimportant, and can be neglected. It is also well known that the
external fields ($\vec{B}_{\infty}$) decrease rather rapidly with distance,
especially above strong field regions such as active regions. For instance, Kliem \& T\"or\"ok (2006) 
have shown that the torus instability, which is one means of accelerating a flux rope CME in its initial stages, is operational only if the external magnetic field falls off faster than $R^{-3/2}$, where $R$ is the distance from the sun center. The external
field will certainly be negligible at distances as large as 2 $R_{\odot}$, which is the starting radius for our observations. This means that the third term on the RHS of Eq~\ref{eq1} can be neglected as well, and it is only the first term that is important. We therefore confine our attention only to this term from now on.

The expression for the Lorentz self-force (the first term on the RHS of Eq~\ref{eq1}) was given by Shafranov (1966), and has since been extensively used by several authors
(e.g., Anzer 1978; Garren \& Chen 1994; Chen 1996; Kumar \& Rust 1996; Titov \& D\'emoulin 1999; Isenberg \& Forbes 2007).
Several of these papers use only the Lorentz self-force term owing to an
implicit recognition that it is by far the most relevant one, and some of
them neglect the rest of the terms in Eq~\ref{eq1} in order to adopt a
largely analytical treatment. We, on the other hand, have clearly demonstrated our rationale in neglecting all the terms save for the one on
the Lorentz self-force, especially for the sample we have chosen.
In this term, the quantity $I_{0}$ is the axial
current, $c$ is the speed of light, $a$ is the major radius of curvature of the curved poloidal flux tube and $r_{0}$ is its minor radius. The quantity $l_{i}$ is of order unity, and denotes the internal inductance per unit length. It depends upon the geometry of current distribution inside the flux rope. For the sake of concreteness, we assume the flux-rope to be characterized by a constant $\alpha$, force-free Lundquist solution in cylindrical geometry (Lundquist 1950; Burlaga et al 1981; DeVore 2000)
{\begin{eqnarray}
\nonumber
B_{r} = 0\, ,\\
\nonumber
B_{\phi} = \sigma_{H}\,B_{0}\,J_{1}\biggl (\alpha_{0} \frac{r}{r_{0}} \biggr ) \, , \\
B_{z} = B_{0}\,J_{0}\biggl (\alpha_{0} \frac{r}{r_{0}} \biggr )\, ,
\label{eq1a}
\end{eqnarray}}
where $B_{0}$ is signed, axial field of the flux rope, $\sigma_{H} = \pm 1$ selects the appropriate handedness near the axis, $r_{0}$ is the flux rope radius, $J_{n}$ is the Bessel function of the first kind of order $n$ and $\alpha_{0}$ is a dimensionless parameter determined by the boundary conditions. Placing the outer boundary of the flux rope at the first zero-crossing of the function $J_{0}$ yields $\alpha_{0} = 2.4$ (DeVore 2000). The internal inductance per unit length ($l_{i}$) is given by
\begin{equation}
l_{i} = 2\, \frac{\int_{0}^{r_{0}} r\, B_{\phi}^{2} (r) dr}{r_{0}^{2} B_{\phi}^{2}(r_{0})} \, .
\label{eq1b}
\end{equation}
Using Eq~\ref{eq1a} in Eq~\ref{eq1b}, we obtain $l_{i} = 1$.

Using these arguments, the force per unit length on the curved portion of the flux rope can be written as 

\begin{equation}
f \, = \frac{I_{0}^{2}}{a \, c^{2}}\,\biggl [ {\rm ln}\biggl (\frac{8\,a}{r_{0}} \biggr ) - 1 \biggr ] \, .
\label{eq2}
\end{equation}

It may be noted that 
this expression is strictly correct only if  the torus is slender; i.e., if the ratio within the logarithm $8a/r_{0} \gg 1$ (Landau, Lifshitz \& Pitaevskii 1984); the relative error is of the order of $[\ln (8a/r_{0})]^{-1}$. The average value for
the ratio $8a/r_{0}$ is $\gtrsim 20$ for the events we consider (column 3, table 1).


\subsection{Calculating the axial current}
The procedure adopted in deriving the mass images and the mechanical energy for each timestamp
for a given CME is described in Vourlidas et al (2000) and SV07.
We circumscribe the extent of the CME cross-section (which is envisaged to be the cross-section
of a magnetic flux rope) at each timestamp. Among other quantities, we measure the position of
the center of mass of the flux rope at each timestamp and the number of pixels enclosed by the
flux rope cross-section. 
We take $a$ to be equal to the height of the CME center of mass.
The total number of pixels enclosed by the CME cross-section gives its area; we derive the
quantity $r_{0}$ from this area by assuming the cross-section to be circular.

For each CME in group A of SV07, 
we compute the driving power by fitting a straight line to the plots of mechanical energy vs. time. An example of an event where the CME is clearly driven (i.e., the mechanical energy increases as a function of time) is shown in figure 1.

\begin{figure*}
\includegraphics[width=3.7in]{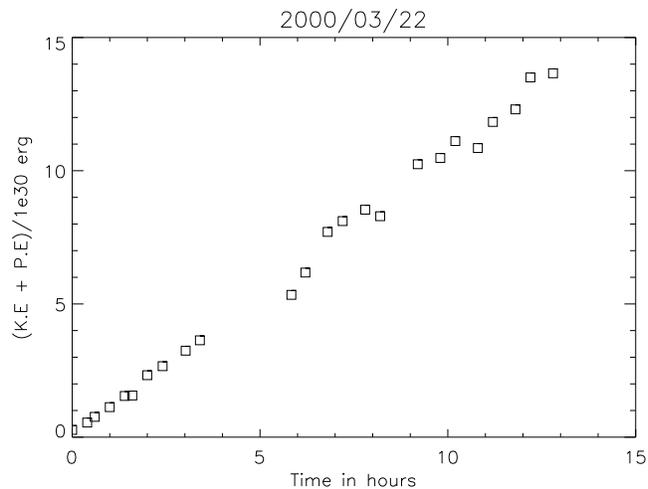}
\caption{Mechanical (i.e., kinetic + potential) energy (in units of
  $10^{30}$ ergs) as a function of time for the flux rope CME of
  2000/03/22}
\end{figure*}

The slope of the straight line fitted to such a plot gives the driving power $P_{D}$. The driving force $F_{\rm D}$ is then computed by dividing
the driving power by the velocity of the CME center of mass. The driving 
force $F_{\rm D}$ is equated to the total Lorentz force
\begin{equation}
F_{\rm L} = \pi a \, f \, ,
\label{eq2a}
\end{equation}
where $f$ is given by Eq~\ref{eq2}. The factor $\pi \, a$ implies a length of $\pi$ radians for the curved portion of the flux rope into the plane of the sky.

Since the driving force $F_{\rm D}$ is derived from a linear fit to the mechanical energy vs. time
profile, we have a single number for this quantity for each CME we have considered in our sample. On the other
hand, we have information for the quantities $a$ and $r_{0}$ for each timestamp for a given CME.
Upon equating $F_{\rm L}$ at each timestamp to $F_{\rm D}$, the only unknown quantity is the axial
current $I_{0}$. For a given CME in our sample, we therefore obtain a value for the axial current $I$ for each timestamp. Examples for some of the CMEs in our sample are shown in figure 2.
\begin{figure*}
\includegraphics[width=3.7in]{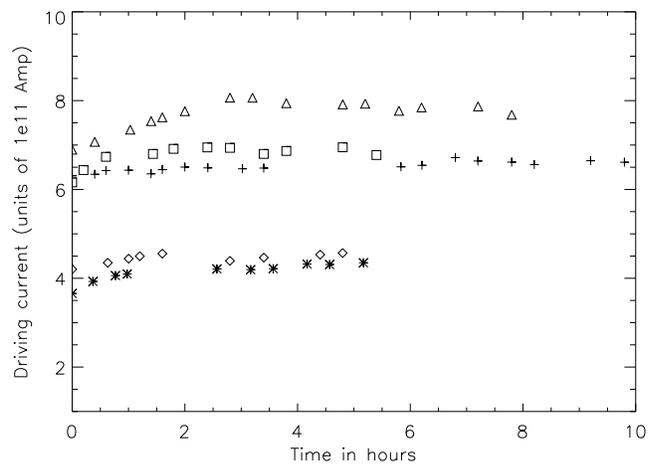}
\caption{Inferred driving current $I_{0}$ for some of the flux rope
  CMEs in our sample. $+$ 2000/03/22, * 2000/06/08, $\Diamond$
  2000/07/23, $\triangle$ 2000/08/02, $\Box$ 2000/08/03} 
\end{figure*}
There is very little variation in the computed driving current $I_{0}$ with time. The small variations
observed are due to changes in the quantities $a$ and $r_{0}$, which might well be due
to projection effects. As mentioned earlier, we have selected CMEs that retain a clear flux rope morphology throughout the field of view, and the projection effects are therefore expected to be small.
The most meaningful way of proceeding is to compute the
average value of the current $\langle I_{0} \rangle$ from plots such as those shown in 
figure 2. The results for all the CMEs in group A of SV07 (i.e., the ones that are clearly driven) are shown in table 1. Evidently, the axial current for the CMEs we have studied here is around a few $\times 10^{10}$ Amperes.

\section{Discussion}

A few caveats are in order; the Lorentz self-force acts only on the topmost, curved part of the flux rope CME. Our calculations cannot distinguish between a flux rope that is completely detached from the solar surface and one that is line-tied and distending outwards in the manner of an aneurism (e.g., Isenberg \& Forbes 2007). The only quantity of interest (in this regard) that enters our calculation is the curved extent of the flux rope; we have uniformly assumed it to be $\pi$ radians (Eq~\ref{eq2a}). It should also be kept in
mind that the fact that we are appealing to a Lorentz self-force means that the flux rope is not force-free. In fact, the manner in which the expression for the Lorentz self-force is derived makes no appeal to misaligned currents and magnetic fields; it is calculated using the spatial gradient of the self-inductance of the flux rope. However, the flux-rope can be almost force-free; the axial current only need be slightly misaligned in order for the Lorentz self-force to be effective (Kumar \& Rust 1996). Furthermore, we have not accounted for drag forces
(e.g., Chen 1996). By equating the 
observationally determined driving force only to the Lorentz self-force (and neglecting drag forces), we are overestimating the Lorentz self-force, and the axial current we compute is therefore an upper limit.

Having estimated the axial current, a useful quantity to estimate is the magnetic flux $\phi$ carried by these CMEs. It is related to their helicity, which some authors believe is a crucial determinant of
their capacity to erupt (e.g., Low 1996; Rust 2001; Nindos et al 2003; Kusano et al 2004). However, there are others (e.g., Phillips, MacNiece \& Antiochos 2005) who argue that there need not be
a critical helicity buildup for CMEs to erupt. Furthermore, the magnetic flux is a quantity that is more easily defined. We therefore estimate it 
at the first timestamp for each of the CMEs in our sample, using Eq (52) of DeVore (2000), which we reproduce below:

\begin{equation}
\phi = 1.4 \, B_{0} r_{0}^{2} \, ,
\label{eq3}
\end{equation} 

where the quantity $B_{0}$ is defined in Eq~\ref{eq1a} and $r_{0}$ is the 
radius of the flux rope as defined in \S~2.2. We relate the axial current
$I_{0}$ to $B_{0}$ using

\begin{equation}
B_{\phi}(r = r_{0}) = \frac{2 I_{0}}{c r_{0}}
\label{eq4}
\end{equation}

Using Eqs~\ref{eq1a}, \ref{eq4} and \ref{eq3}, we get

\begin{equation}
\phi = \frac{2.8 I_{0} r_{0}}{\sigma_{H} c J_{1} (\alpha_{0})}
\label{eq5}
\end{equation}

The quantity $\phi$ for each of the CMEs we have studied is quoted in column 5 of Table 1. 
\begin{table*}
\caption{Average driving current enclosed by CMEs}
\label{table:1}     
\centering 
\begin{tabular}{c c c c c}        
\hline\hline                 
Date & Time & $8 a/r_{0}$ & $\langle I_{0} \rangle$/($10^{10}$ Amp) & $\phi/(10^{21} {\rm Mx})$\\
\hline 
97/11/01  & 20:11 & 21 & 1 & 2\\
97/11/16  & 23:27$^{\mathrm{a}}$ & 19 & 1 & 10\\ 
98/02/04  & 17:02 & 18 & 2 & 6\\
98/02/24  & 07:28 & 26 & 0.5 & 2\\
98/05/07  & 11:05 & 22 & 2 & 7\\
98/06/02  & 08:08 & 14 & 3 & 10\\
99/07/02  & 17:30 & 14 & 1 & 7\\
99/08/02  & 22:26 & 26 & 1 & 4\\
00/03/22  & 04:06 & 18 & 2 & 6\\
00/05/05  & 07:26 & 23 & 1 & 2\\
00/05/29  & 04:30 & 20 & 1 & 5\\
00/06/06  & 04:54 & 21 & 1 & 5\\
00/06/08  & 17:07 & 13 & 1 & 3\\
00/07/23  & 17:30 & 16 & 1 & 4\\
00/08/02  & 17:54 & 14 & 2 & 5\\
00/08/03  & 08:30 & 15 & 2 & 5\\
00/09/27  & 00:50 & 15 & 1 & 4.2\\
00/10/26  & 00:50 & 17 & 1 & 4\\
00/11/12  & 09:06 & 22 & 1 & 11\\
00/11/14  & 16:06 & 19 & 1 & 1\\
00/11/17  & 04:06 & 20 & 1 & 2\\
00/11/17  & 06:30 & 18 & 1 & 3\\
01/01/07  & 04:06 & 14 & 1 & 3\\
01/01/19  & 17:06 & 18 & 1 & 5\\
01/02/10  & 23:06$^{\mathrm{a}}$ & 11 & 2 & 19\\
01/03/01  & 04:06 & 14 & 1 & 2\\
01/03/23  & 12:06 & 15 & 2 & 5\\
\hline                                   
\end{tabular}
\begin{list}{}{}
\item[$^{\mathrm{a}}$] The time refers to the previous day.
\item[] \textsl{Column 1\/}: Date on which a given CME
  occurred; \textsl{Column 2\/}: Start time in the C2 field of view;
\textsl{Column 3\/}: The quantity $a/r_{0}$ (Eq~\ref{eq1});
  \textsl{Column 4\/}: Average driving current in units of $10^{10}$ Amperes; \textsl{Column 5\/}: Magnetic flux at starting timestamp in units of $10^{21}$ Mx.
\end{list}
\end{table*}
We note that 
$\phi$ is generally a few times $10^{21}$ Mx. This compares well with the generally quoted value of $10^{21}$ Mx for the 
average flux carried by near-earth magnetic clouds (e.g., Lepping, Jones \& Burlaga 1990; DeVore 2000). Since we are invoking dissipation of magnetic flux via Lorentz self-forces in order to
explain the driving force on CMEs, it is understandable that the flux
carried by a typical CME towards the start of its journey is somewhat larger than
what it carries when it reaches the earth in the form of a magnetic cloud. 

\section{Conclusions}
We have computed upper limits on the axial currents enclosed by
flux-rope CMEs. Our method relies on measurements of the driving
power for these CMEs using a well established method using
LASCO data. We assume that the driving force is entirely due to
Lorentz self-forces in the bent torus comprising the flux rope.
We have chosen a sample of flux rope CMEs that clearly experience a
driving force in the LASCO field of view.
We find that the average driving current for each of the CMEs in our sample
is a few $\times 10^{10}$ Amperes.
This figure is about an order of magnitude lower than estimates
of currents carried by filaments (e.g., Ballester \& Kleckzek
1984). Estimates of currents in active region flaring loops range from
$10^{10}$--$10^{12}$ Amperes (e.g., Zaitsev et al 1998; Tan et al
2006). On the other hand, using a method that involves measuring the polarization of
radio sources observed against the solar corona, Spangler (2007) has
estimated coronal currents ranging from $10^{8}$--$10^{9}$ Amperes. It may be emphasized, however, that Spangler's measurement pertained to the current enclosed in an Amperian loop in the quiescent solar corona, and had nothing to do with CMEs or their driving currents. As mentioned in \S~1, the only attempt to apply the Faraday rotation technique to CMEs has succeeded only in determining the magnetic field orientation, and not its magnitude.

We also note that we obtain values of a few $\times 10^{21}$ Mx for the flux carried by the CMEs at the first
timestamp (i.e., towards the beginning of their journey). This value is a factor of a few
larger than the generally quoted average value of $10^{21}$ Mx for the flux carried by an average near-earth magnetic cloud. The excess flux is presumably dissipated (via Lorentz self-forces) in powering the CME during its journey from the sun to the earth. This fits in very well with our overall picture of CME energetics (Vourlidas et al 2000; SV07). Given the completely different data sources used to estimate the flux carried by near-earth magnetic clouds (Lepping, Jones \& Burlaga 1990; DeVore 2000) and that carried by flux rope CMEs (this work), this level of agreement is remarkable, and lends strong support to our overall hypothesis.

Finally, we comment on the utility of our results in the light of the numerous analytical and numerical attempts at describing CME energetics. Our approach has been to concentrate on events for which there is clear evidence of driving power. Furthermore, the driving power for these events is reasonably
constant in the field of view, as evident from the largely linear shape of the mechanical energy vs time plots. As explained in \S~2.2, this results in a single number for the driving force, and consequently the driving current, throughout the field of view. In some sense, our results should be compared with the regime in the simulations which show evidence for a constant driving current. The most appropriate example we could find was Figure 4 of Isenberg \& Forbes (2007). The semi-analytical model described there predicts a significant range where the driving current approaches a constant value. However, it may be noted that such models (as well as simulations) can only predict the shape of the drive current vs time curve, and cannot assign a number to the normalization. Our results are complementary in the sense that they provide a definite number (a few $\times 10^{10}$ Amperes) for the asymptotic value of the drive current in such a model, thereby fixing the normalization.

We plan to extend our measurements to CME observations from the SECCHI coronagraphs aboard the STEREO mission. These observations can provide a much better estimate of the three-dimensional extent of a CME and will improve the accuracy of our current estimates.

\section{Acknowledgements}
We thank the anonymous referee for a critical appraisal of our work that has helped us significantly improve this paper.

\end{document}